\pdfoutput=1
\documentclass[fleqn,usenatbib]{mnras}

\usepackage{newtxtext,newtxmath}

\usepackage[T1]{fontenc}
\usepackage{ae,aecompl}
\usepackage{soul}

\usepackage{graphicx}	
\usepackage{amsmath}	
\usepackage{amssymb}	
\usepackage{enumerate}

\title[Radio Galaxy Detection in the Visibility Domain]{Radio Galaxy Detection in the Visibility Domain}

\author[A. Malyali et al.]{
A. Malyali,$^{1, 2}$\thanks{E-mail: amalyali@mpe.mpg.de}
M. Rivi,$^{1,3}$
F. B. Abdalla,$^{1,4§}$
J. D. McEwen$^{5}$
\\
$^{1}$Department of Physics and Astronomy, University College London, Gower Street, London, WC1E 6BT, U.K.\\
$^{2}$Max-Planck-Institut f\"ur extraterrestrische Physik, Giessenbachstrasse 1, 85748, Garching, Germany \\
$^{3}$INAF - Istituto di Radioastronomia, via Gobetti 101, 40129 Bologna, Italy\\
$^{4}$Department of Physics and Electronics, Rhodes University, PO Box 94, Grahamstown, 6140, South Africa\\
$^{5}$Mullard Space Science Laboratory, University College London, Surrey RH5 6NT, U.K.
}

\date{Accepted XXX. Received YYY; in original form ZZZ}

\pubyear{2018}

\begin{document}
\label{firstpage}
\pagerange{\pageref{firstpage}--\pageref{lastpage}}
\maketitle

\begin{abstract}
We explore a new Bayesian method of detecting galaxies from radio interferometric data of the faint sky. Working in the Fourier domain, we fit a single, parameterised galaxy model to simulated visibility data of star-forming galaxies. The resulting multimodal posterior distribution is then sampled using a multimodal nested sampling algorithm such as \textsc{MultiNest}. For each galaxy, we construct parameter estimates for the position, flux, scale-length and ellipticities from the posterior samples. We first test our approach on simulated SKA1-MID visibility data of up to 100 galaxies in the field of view, considering a typical weak lensing survey regime (SNR~$\ge 10$) where 98\% of the input galaxies are detected with no spurious source detections. We then explore the low SNR regime, finding our approach reliable in galaxy detection and providing in particular high accuracy in positional estimates down to SNR~$\sim 5$.
The presented method does not require transformation of visibilities to the image domain, and requires no prior knowledge of the number of galaxies in the field of view, thus could become a useful tool for constructing accurate radio galaxy catalogs in the future.
\end{abstract}




\section{Introduction}
The detection of galaxies from radio interferometric data has traditionally relied on a set of image reconstruction techniques to transform measured visibilities to the image domain \citep{hogbom1974aperture,Bhatnagar2004,cornwell2008multiscale}, on which image analysis tools are applied to measure source properties. 
The imaging process usually transforms visibilities into images, corrected for an approximated Point Spread Function, by an iterative non-linear process. This can introduce artifacts and correlated noise distributions, with subsequent measurements of scientific parameters suffering from systematic errors that are difficult to accurately estimate. In addition, uncertainties in the image domain are typically not accessible, although recent work can provide uncertainty quantification, e.g. \cite{cai2018uncertainty, cai2018uncertainty2}. 
 
Until recently this has not been a major issue, but the increased sensitivities and size of the forthcoming generation of radio interferometers, such as the Square Kilometre Array (SKA)\footnote{\url{https://skatelescope.org/}} \citep{dewdney2009square}, will allow new scientific measurements, such as radio weak lensing \citep{Brown2015}, that require more reliable and complete source catalogs, meaning higher accuracy in galaxy detection and characterization. Observations of large fields of view, or with non-coplanar baselines, will be also affected by direction dependent effects (DDE) that should be taken into account in the data analysis \citep{wrobel1999sensitivity}.
An alternative approach to usage of image reconstruction techniques is to work directly in the visibility domain, where the data originates and is not yet affected by the systematics introduced by the imaging process. DDE modeling, e.g. \citet{smirnov2011revisiting2,Smirnov2011b}, may also be easily introduced in model fitting techniques for galaxy parameter estimation. However, adopting a source model fitting approach to visibilities is very challenging because sources are no longer localized in the Fourier domain and their flux is mixed together in a complicated way. 
Furthermore the computational challenges for a telescope such as SKA are great, given the large size of datasets (order of PetaBytes) that must be processed and the expected high source number densities. 

Available generic tools for model fitting in the Fourier domain are based on simple source brightness profiles or models obtained from the combination of basic shapes that are not sufficiently realistic \citep{MartiVidal2014}. Other methods are specific for galaxy shape measurement for radio weak lensing and rely on information, such as the number of sources and galaxy positions,
determined via some preprocessing of the dataset or data analysis in the image domain. The first one decomposes galaxy shapes through an orthonormal basis of functions, \textit{shapelets} \citep{Refregier2001,Refregier2001a}, and coefficients are jointly fitted solving normal equations \citep{Chang2004}. The second one, \textit{RadioLensfit} \citep{rivi2016radiolensfit,rivi2018radio}, is an adaptation to the radio data of the optical method {\it lensfit} \citep{miller2013bayesian}, where a source extraction technique has been implemented in order to fit visibilities of a single galaxy at a time. The single source ellipticity fitting is performed adopting a realistic star-forming galaxy brightness profile and marginalising the likelihood over flux, position and size. 
Another approach \citep{rivi2018radio2} applies Hamiltonian Monte Carlo (HMC) technique for a joint fitting of all sources in the field of view adopting the same galaxy model used for \textit{RadioLensfit}. Although HMC accelerates posterior sampling convergence with respect to standard Monte Carlo Markov Chain (MCMC) with a reasonable efficiency even for high dimensional problems, it is still a computationally demanding approach to apply to the very large number of sources that will be observed in future radio surveys. Moreover, if the number of sources in the dataset is unknown, then the posterior sampling algorithm must be able to handle changing dimensions of the parameter space, such as via reversible-jump MCMC \citep{green1995reversible} for the case of a multiple source model.
A single object model approach can avoid this changing dimensionality issue due to forming a multimodal posterior in the single object parameter space, with the bonus that it is typically computationally cheaper as the number of parameters to fit are limited to the ones of a single source.

A first attempt to detect discrete objects in astronomical datasets with a single object model is presented in \cite{hobson2003bayesian}, where two iterative methods (using MCMC and a simulated annealing simplex technique) are applied to microwave maps dominated by emission from primordial cosmic microwave background anisotropies  in order to detect the thermal and kinetic Sunyaev-Zel'dovich effects from clusters of galaxies. 
\cite{Feroz2008cluster} also demonstrate the strength of the single model approach using multimodal nested sampling to detect galaxy clusters and estimate their parameters from images of N-body simulations.

In this work, we explore a similar Bayesian technique for detecting and characterising star-forming galaxies in the Fourier domain. We fit a single galaxy model to simulated interferometer data and use the multimodal nested sampling algorithm \textsc{MultiNest}~\citep{Feroz2008,Feroz2009,feroz2013importance} for detecting multiple galaxy sources in the data and measuring each galaxy's properties. This approach is self-consistent, as it does not require knowledge in advance of the number of galaxies in the field of view.  
We present the relevant theoretical background and our method in Section~\ref{sec:methodology}. We then explore the detection ability of this approach on simulated radio observations in Section~\ref{sec:sim_radio_observations}, and conclude with a discussion and summary of the effectiveness of this new method in Section~\ref{sec:conclusions}.

\section{Methodology}
\label{sec:methodology}
We follow the model fitting approach presented in \cite{Feroz2008cluster} adopting a single galaxy model. This model allows for working with a small number of parameters, but returns a highly multi-modal posterior that is then sampled using \textsc{MultiNest}. Each mode should correspond to a different galaxy in the field of view, however noise and source contaminations usually produce a number of false positives that must be recognised.
We begin by briefly introducing the relevant underlying Bayesian framework in Section~\ref{sec:Bayes} and then present details of the adopted model and parameters' prior distributions in Sections~\ref{sec:likelihood} and \ref{sec:priors}.

\subsection{Bayesian statistics}
\label{sec:Bayes}
Bayesian methods for the estimation of a set of parameters~$\mathbf{\Theta}$ in a model/hypothesis $H$ for given data $\mathbf{D}$ estimate the posterior probability distribution $\mathrm{Pr}(\mathbf{\Theta} \mid \mathbf{D},H)$ of the values of $\mathbf{\Theta}$. Bayes' theorem reads:
\begin{equation} \label{eqn:bayes_theorem_data}
\mathrm{Pr}(\mathbf{\Theta} \mid \mathbf{D},H) = \frac{\mathrm{Pr}(\mathbf{D} \mid \mathbf{\Theta},H) \, \mathrm{Pr}(\mathbf{\Theta} \mid H)}{\mathrm{Pr}(\mathbf{D} \mid H)},
\end{equation}
where the likelihood $\mathcal{L}(\mathbf{\Theta}) \equiv \mathrm{Pr}(\mathbf{D} \mid \mathbf{\Theta},H)$ encodes the constraints imposed by observations, $\pi (\mathbf{\Theta}) \equiv \mathrm{Pr}(\mathbf{\Theta},H)$ is the probability distribution of the parameters based on prior knowledge of the parameter values (prior), and $\mathcal{Z} \equiv \mathrm{Pr}(\mathbf{D} \mid H)$ is a normalisation factor, known as the {\it Bayesian evidence}, corresponding to the average of the likelihood weighted by the prior over its $N$-dimensional parameter space:
\begin{equation} \label{eqn:evidence}
\mathcal{Z} = \int \mathcal{L}(\mathbf{\Theta}) \pi(\mathbf{\Theta}) \mathrm{d}^N\mathbf{\Theta}.
\end{equation} 
Parameter estimates along with their associated uncertainties are derived by sampling the un-normalised posterior, as the evidence is independent of the parameters~$\mathbf{\Theta}$. 
$\mathcal{Z}$ becomes relevant for model selection, where a choice between two models $H_0$ and $H_1$ has to be decided based on the comparison of their respective posterior probabilities given the observed data $\mathbf{D}$.
Applying Bayes' theorem to invert the order of conditioning in the evidence, we obtain the model posterior probability:
\begin{equation} 
\mathrm{Pr}(H_k\mid\mathbf{D}) = \frac{\mathrm{Pr}(\mathbf{D}\mid H_k)\mathrm{Pr}(H_k)}{\mathrm{Pr}(\mathbf{D})}, \quad k=0,1.
\end{equation}
Then, the ratio between the models' posteriors is:
\begin{equation}\label{eqn:bayes_factor}
R=\frac{\mathrm{Pr}(H_1\mid\mathbf{D})}{\mathrm{Pr}(H_0\mid\mathbf{D})}=\frac{\mathcal{Z}_1}{\mathcal{Z}_0}\frac{\mathrm{Pr}(H_1)}{\mathrm{Pr}(H_0)}.
\end{equation}
The ratio of the models' evidences is called the \textit{Bayes factor} and reflects the relative strength of support for each model given the data.
More details about Bayesian inference in the context of astrophysics are provided in \cite{trotta2008bayes}.

{\it Nested sampling} \citep{skilling2004nested,skilling2006nested} is a method mainly used for efficient and accurate evidence evaluation relative to traditional MCMC methods, yet also provides draws from the posterior as a by-product. The boost in efficiency arises from converting a multi-dimensional integral into effectively a one-dimensional integral, achieved through the relation between enclosed prior mass, $X(\lambda)$, 
and likelihood:
\begin{equation} \label{eqn:prior_volume}
X(\lambda) = \int_{\mathcal{L}(\mathbf{\Theta}) > \lambda} \pi(\mathbf{\Theta})\mathrm{d}^N\mathbf{\Theta}.
\end{equation}
The nested sampling algorithm begins with drawing a set of independent random points from the prior (\textit{live points}). In each iteration, the point with lowest likelihood, $\mathcal{L}_\mathrm{old}$, is removed and replaced with a new point randomly drawn from the prior, subject to the constraint that its likelihood $\mathcal{L}$ is greater than $\mathcal{L}_\mathrm{old}$. 
This process continues until convergence is reached; typically when the fraction of live enclosed posterior mass is a small fraction of the overall mass \citep{keeton2011statistical}. The evidence is then obtained through quadrature:
\begin{equation} \label{eqn:ns_approx_ev}
\mathcal{Z} \approx \sum_{i=1}^{N_p} \mathcal{L}_iw_i = \frac{1}{2} \sum_{i=1}^{N_p} \mathcal{L}_i(X_{i-1}-X_i),
\end{equation}
where $N_p$ is the number of discarded points, $\mathcal{L}_i$ is the likelihood of the $i$-th discarded point, $w_i=(X_{i-1}-X_i)/2$ is the weight assigned to each point. Draws from the posterior, $p_i$, are obtained as:
\begin{equation} \label{eqn:ns_posterior_draws}
p_i = \frac{\mathcal{L}_i w_i}{\mathcal{Z}}
\end{equation}
and these samples can then be used to estimate the mean and standard deviation of the model parameters. 

\textsc{MultiNest}\footnote{\url{https://ccpforge.cse.rl.ac.uk/gf/project/multinest/}}  is an implementation of nested sampling based on ellipsoidal rejection sampling and tailored for sampling multimodal posteriors. It identifies peaks in the posterior distribution and returns a set of posterior samples for each mode. \textsc{MultiNest} also returns the \textit{local evidence} $\mathcal{Z}_\mathrm{loc}$ of each mode based on the set of posterior samples associated with it (as defined in Section 5.7 of \citealt{Feroz2009}). Since we use only a six parameter model, \textsc{MultiNest} has the best performance amongst other available nested sampling implementations, such as \textsc{PolyChord} \citep{handley2015polychord,handley2015polychord2}, which scale well to higher dimensional settings. 

\subsection{Galaxy likelihood}
\label{sec:likelihood}
We adopt a single star-forming galaxy visibility model defined analytically as the Fourier transform of a S\'ersic model of index $n=1$ (exponential), to estimate a galaxy's position $(l,m)$, flux~$S$, scale-length~$\alpha$ and ellipticity components $(e_1, e_2)$ \citep{rivi2016radiolensfit}. This galaxy model is defined as only having a disk component because radio emission is dominated by synchrotron radiation from relativistic electrons in the interstellar medium of the galaxy disk. 

Following \cite{rivi2018radio2}, model visibilities are simulated by using the GPU-accelerated implementation of the radio interferometer measurement equation \citep{hamaker1996understanding, smirnov2011revisiting, smirnov2011revisiting2}, provided by the open-source software \textsc{Montblanc}\footnote{\url{https://github.com/ska-sa/montblanc/}} \citep{perkins2015montblanc1}. This tool, developed in support of Bayesian inference of radio observations \citep{BIRO}, also returns the likelihood computation given the observational data and the noise variance. 

\subsection{Priors}
\label{sec:priors}
 
We adopt a uniform prior on source position over the interferometer's field of view (FOV). For the other galaxy parameters we use distributions presented in \cite{rivi2018radio} and estimated from measurements of faint sources observed with the Very Large Array (VLA) radio telescope. In particular the flux and scale-length priors are obtained by the analysis of the deep radio VLA-SWIRE field catalog:
\begin{equation}
\label{eq:flux}
\pi(S) \propto S^{-1.34}
\end{equation}
and $\pi(\alpha)$ is a log-normal distribution (flux independent) with mean $\mu \sim 0.27$ arcsec and variance $\sigma \sim 0.31$ arcsec.
The prior on the ellipticity modulus is obtained by fitting the function proposed in \citet{miller2013bayesian} to the VLA-COSMOS field data, using a maximum ellipticity value of 0.804 as in the optical case.

\section{Testing on SKA1-MID simulated radio observations}
\label{sec:sim_radio_observations}

\subsection{Data simulation}
\label{sec:sim}  

SKA-MID will be a dish array located in South Africa made, in Phase~1, of 64 MeerKAT dishes in a moderately compact core with a diameter of about 1~km and 133 SKA1 dishes distributed in the core and in three logarithmically spaced spiral arms emanating from the centre and extending out to a maximum radius of 80~km, with a maximum baseline of 150~km. 
We use the antennae configuration provided in \citet{heystek2015ska} to simulate 8 hour observations of a 1 deg$^2$ FOV at declination of~$-30^{\circ}$ (i.e. at the zenith), in a single smeared-out frequency channel between 1280-1520 MHz and with visibilities sampled once every $\tau _{\mathrm{acc}}=60$~s. 

We generate realistic star-forming galaxy populations as in \citet{rivi2018radio}, where sources are randomly distributed within the FOV and scale-lengths are flux dependent according to the following linear relation between the log median values: $\ln{[\alpha_\textrm{med}/\textrm{arcsec}]} = -0.93 +0.33\ln{[S/\mu \textrm{Jy}]}.$
Observed visibilities are computed adopting the exponential profile, as for the model visibilities, and we add uncorrelated Gaussian noise 
to these with variance given in \cite{wrobel1999sensitivity} assuming all the antennas are SKA1-MID dish antennae.

\subsection{Source detection results}
\label{sec:modes}
We use the Python implementation of \textsc{MultiNest}, PyMultiNest\footnote{\url{https://johannesbuchner.github.io/PyMultiNest}} \citep{Buchner2014AstrophysicsCatalogue} for sampling the posterior, running it in multimodal mode and setting the sampling efficiency to 0.8. This corresponds to the enlarging of the ellipsoidal bounds by a factor of 1.25 between iterations.
The number of live points, $N_{\textrm{live}}$, used determines the resolution of the algorithm at finding posterior modes. If $N_{\textrm{live}}$ is too small, then \textsc{MultiNest} can miss drawing samples from a subset of the modes and no information can be obtained about that region of the posterior (see Fig.~\ref{fig:flux_cut_n_clusters} where lower $N_{\textrm{live}}$ values result in fewer detected sources). This could lead to inferences between different seeded runs of \textsc{MultiNest} being inconsistent, due to two different runs potentially having vastly different sampled posteriors, thus it is crucial that $N_{\textrm{live}}$ is set to a sufficiently high value.
On the other side a very large number of live points increases the computational cost and may produce a significant number of fake modes when the likelihood function is not sufficiently smooth. This could happen in the case of too much noise in the data, or due to interference with the signal of neighbourhood sources.
Since the number of posterior modes is dependent on the number of simulated galaxies in the field of view, $N_{\textrm{gal}}$, we vary $N_{\textrm{live}}$ between simulations according to $N_{\textrm{gal}}$\footnote{Based on instrument detection sensitivity and simulations of extragalactic radio source populations, estimates of the expected source number densities may be computed for real data (e.g. see \cite{RedBook2018} for the expected number of galaxies for the two planned continuum surveys with SKA1).}. We observe that the total number of modes detected by \textsc{MultiNest} is typically proportional to the number of galaxies and the number of live points used.
For more accurate object detection, we set clustering on the live points to be performed only on the source position parameters $l$, $m$ within the sampling routine. In these parameter spaces, posterior samples from each mode generally have the least overlap with other modes (effectively non-degenerate in cases of non-overlapping sources), compared with the highly degenerate spaces of scale-length, flux and ellipticity ($\alpha$, $S$, $e_1$, $e_2$). 

\begin{figure}
	\centering
    \includegraphics[scale=0.4]{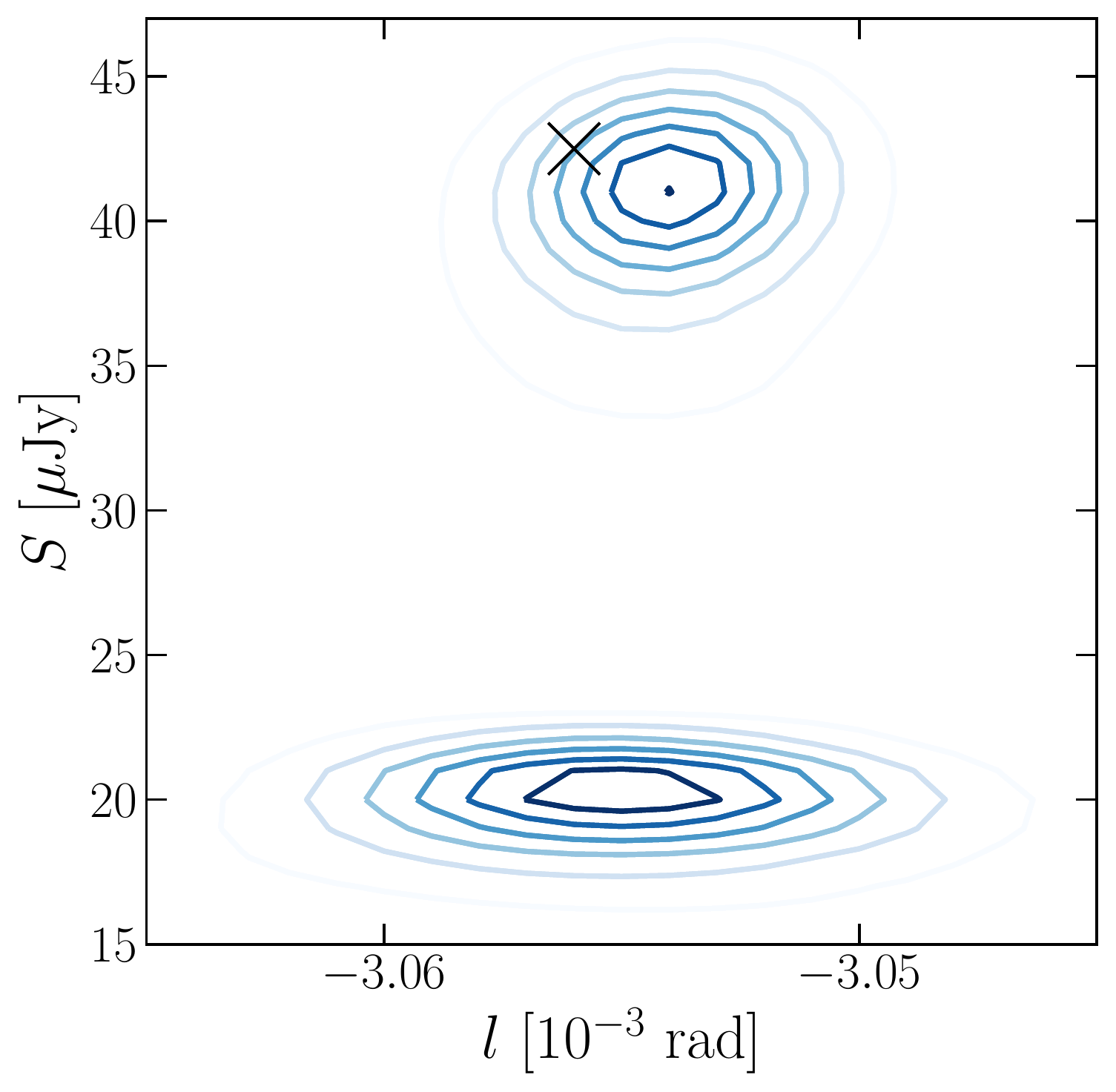}
    \caption{Position $l$ vs flux density $S$ of posterior samples plotted for two modes returned by $\textsc{MultiNest}$. The black cross marks the position and flux of the true galaxy. Only the top mode, which has the highest $\mathcal{Z}_{\textrm{loc}}$ within the cluster, samples from the correct region. The second one is an example of an F1 mode. Coordinate $m$ for each mode is consistent with each other, but it is omitted for illustrative purposes. The darker the contour, the higher the density of posterior samples in that parameter space region.}
  \label{fig:cluster_choosemax_z_local}
\end{figure}

As in \cite{Feroz2008cluster}, we find that the high sensitivity of \textsc{MultiNest} to structure within the posterior results in it returning a larger number of posterior modes than the number of sources. There is typically a population of true modes that correspond to galaxies, and `fake' modes that are spurious. The fake modes can be further classified as two different types, which we call here as F1 and F2 modes. 
F1 modes generally have position estimates that are consistent with a galaxy source, resulting in small clusters of modes forming very close to the locations of true modes\footnote{Despite performing clustering only in positional space, it seems that the default tolerance of \textsc{MultiNest} for splitting live points into separate clusters is too stringent for our application, resulting in multiple modes effectively at the same position. Therefore we need to perform some additional clustering (see Section~\ref{sec:clustering}).}. However, their $\alpha$, $S$, $e_1$, $e_2$ estimates are inconsistent with the corresponding true galaxy parameters - for example, their fluxes are greatly underestimated (see Fig. \ref{fig:cluster_choosemax_z_local}) and they have a lower local evidence~$\mathcal{Z}_{\textrm{loc}}$ than the true mode in the cluster.

\begin{figure}
\centering
    \includegraphics[scale=0.4]{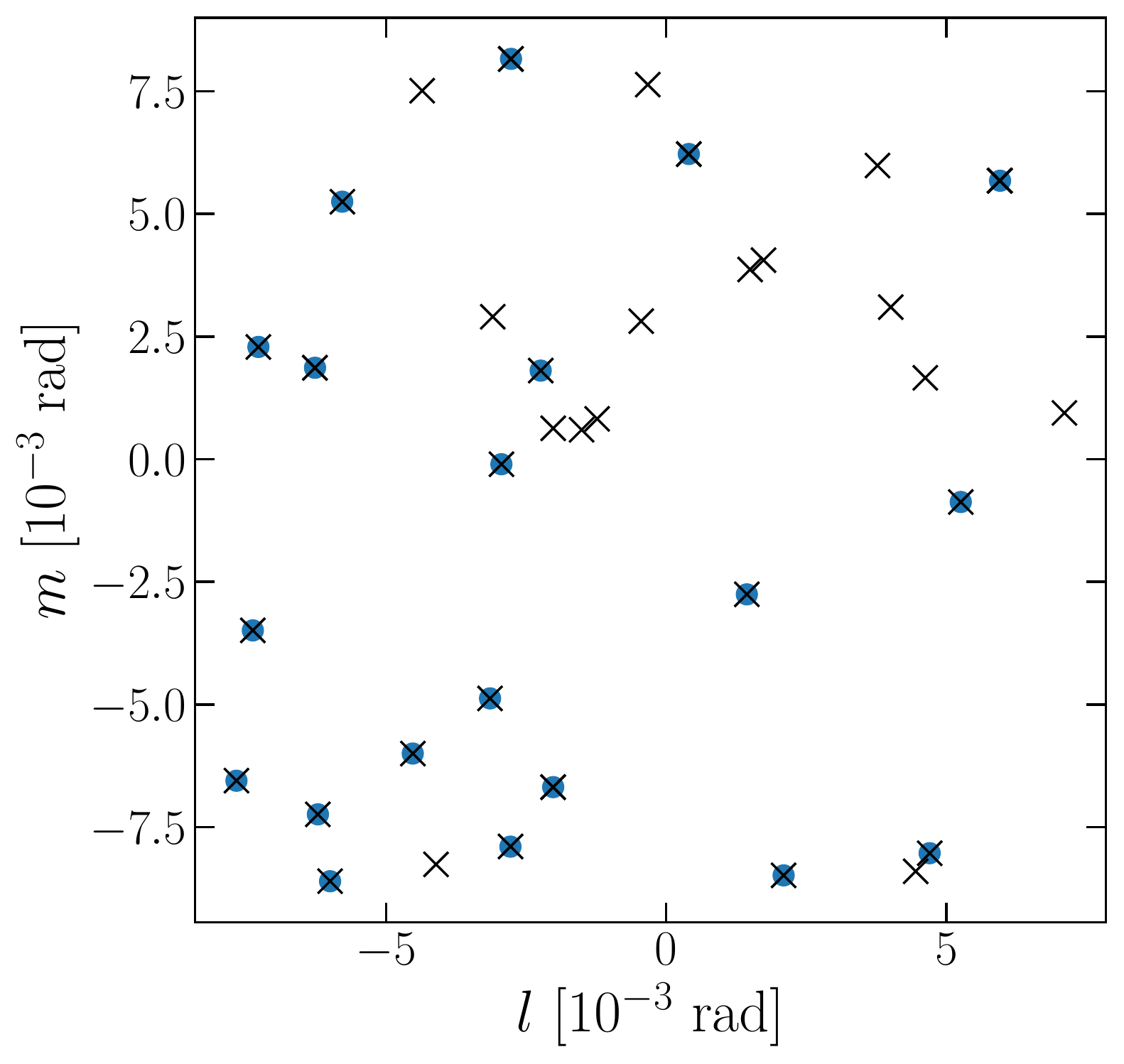}
    \caption{An example of the problem caused by `fake' modes. Blue circles represent the positions of the 20 simulated galaxies, with flux ranging between $10\mu$Jy and $200\mu$Jy, whilst black crosses are the $l$, $m$ estimates for all modes returned by \textsc{MultiNest} (in some cases they are single modes whereas in others they are clusters of modes). The lower bound of the flux prior is 9~$\mu$Jy. F2 modes generally have fluxes between 9-10~$\mu$Jy.}
    \label{fig:spurious_mode_illustration}
\end{figure}
\begin{figure}
\centering
    \includegraphics[scale=0.6]{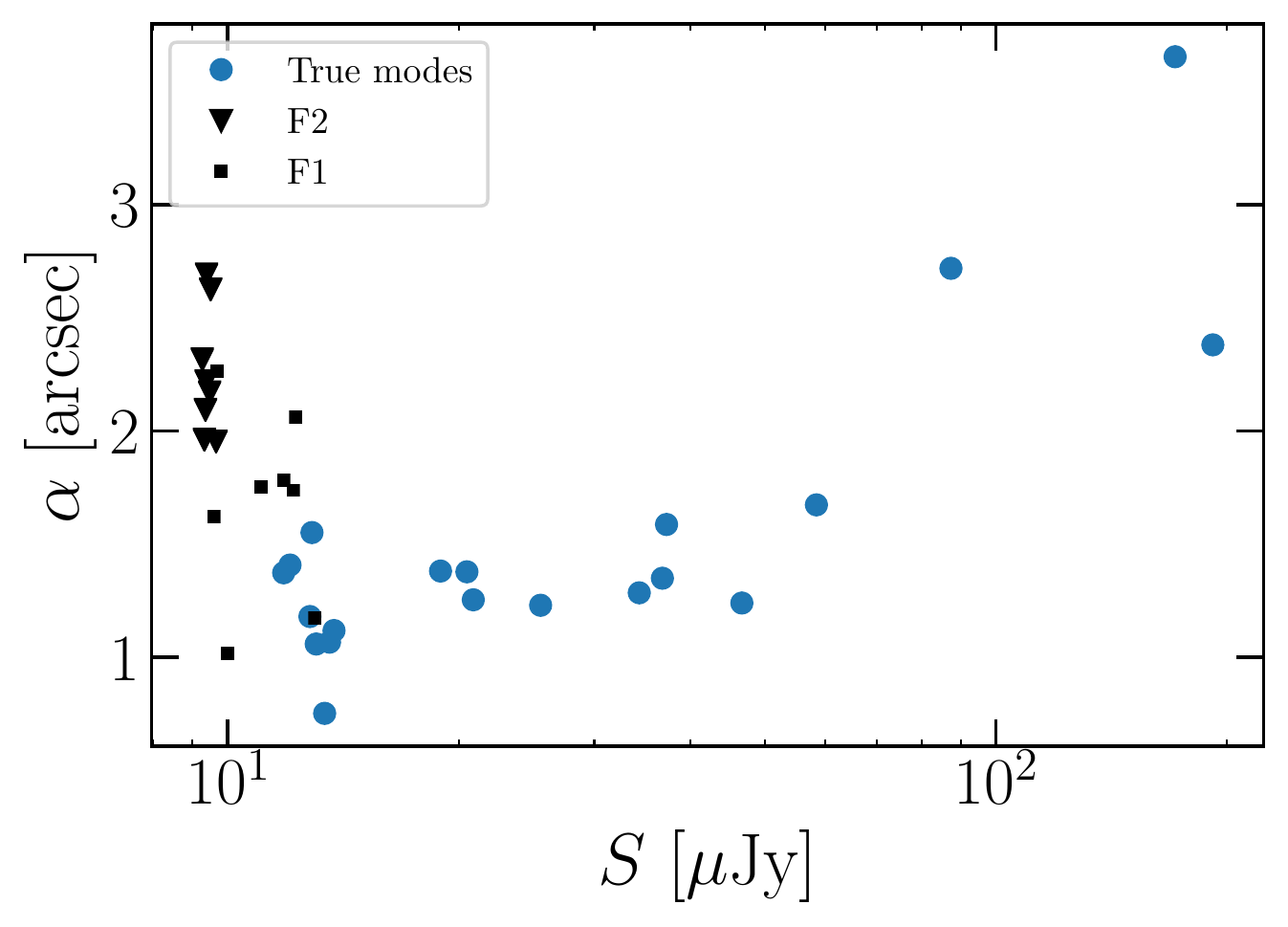}
    \caption{Estimated scale-lengths and fluxes for all modes returned by \textsc{MultiNest} on a simulated population of 20 galaxies, with flux prior ranging between $9\mu$Jy and $200\mu$Jy. Blue circles represent the `true' estimated modes, whilst F1 and F2 modes are shown in black squares and triangles respectively. We see that F2 modes have significantly larger than expected scale-lengths given their low estimated fluxes (relative to the population of true modes).}
    \label{fig:anomalous_scalelengths}
\end{figure}

F2 modes are those which appear at seemingly random positions within the field of view, which vary (in number and also parameter estimates) between different seeded sampling runs. In Fig.~\ref{fig:spurious_mode_illustration} we show an example of modes (plotted as black crosses) detected by \textsc{MultiNest} from a simulated observation of 20 galaxies with flux ranging between $10\mu$Jy and $200\mu$Jy, corresponding to a signal-to-noise ratio SNR~$\ge 10$. F2 modes are the crosses which have no true galaxy counterpart (blue circles), whilst F1 modes generally overlap each other and are located on the blue circles (thus not distinguishable by eye). 
F2 modes typically have a much lower local evidence and their fluxes lie close to the lower bound of the flux prior due to posterior sampling for these modes mainly driven by the shape of the prior. In addition, their estimated scale-lengths are much larger than expected on average from real sources at such low flux (see Fig.~\ref{fig:anomalous_scalelengths}), resulting in the peak brightness and SNR of F2 modes being distinctly low relative to the population of true modes.

\subsection{Modal selection}
If we na\"ively assumed all returned modes were galaxies, then the presence of the fake modes would lead to a high number of false positives that would make accurate galaxy detection unfeasible. In the following, we present a method to remove F1 modes via clustering (see Section~\ref{sec:clustering}) and then we discuss in Section~\ref{sec:spurious} two approaches 
to pick out from the remaining F2 modes the true detections, assuming no prior knowledge about the number of galaxies within the field of view.

\subsubsection{Clustering}
\label{sec:clustering}
Since F1 modes are grouped in clusters around the true galaxy positions, we use an unsupervised, non-parametric clustering algorithm called \textit{mean shift}, implemented in the python package \textsc{scikit-learn}\footnote{\url{http://scikit-learn.org/stable/modules/generated/sklearn.cluster.MeanShift.html}} and run with bandwidth $10^{-4}$ rad (approximately the uncertainty in positional parameter estimates), to identify the centre positions of these mode clusters and the corresponding modes contained in each. For each detected modal cluster, we identify the mode with the highest local evidence and discard the remaining modes. Our final parameter estimate for the galaxy at the cluster centre is then constructed from the posterior samples of the remaining mode. We only select this one, and do not combine the posterior samples from all modes within a cluster to construct parameter estimates for the galaxy, because only the highest $\mathcal{Z}_{\textrm{loc}}$ mode typically samples from a region of the posterior consistent with the true parameter values (see Fig.~\ref{fig:cluster_choosemax_z_local}). This results in only true or spurious source detections.

\subsubsection{Removing spurious detections}
\label{sec:spurious}
As mentioned in Section~\ref{sec:modes}, F2 modes have low SNR and flux estimates heavily dependent on the range of the flux prior provided to \textsc{MultiNest}.
Due to this, a natural choice for removing them is to choose a suitable lower bound for the flux prior and discarding all modes below a SNR threshold.   

\begin{figure}
	\centering
    \includegraphics[scale=0.48]{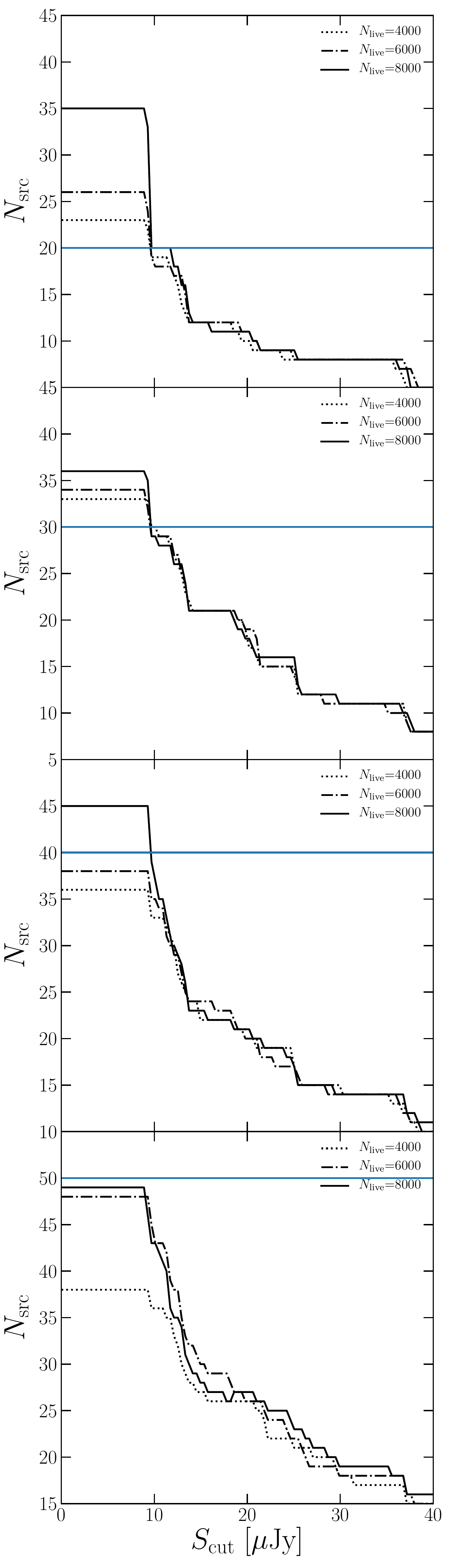}
    \caption{The number of clustered modes, each corresponding to a single source detection, remaining after discarding modes below different flux cuts. From top to bottom, results refer to $N_\mathrm{gal}$ = 20, 30, 40, 50, with blue line representing $N_\mathrm{gal}$ in each plot.  For each case we show plots corresponding to three different numbers of live points: $N_\mathrm{live}$ = 4000, 6000, 8000. The simulated sources have flux $S \ge 10\mu$Jy and flux prior down to $9\mu$Jy.}
    \label{fig:flux_cut_n_clusters}
\end{figure}

Since this approach could be too empirical, we compare it with a more rigorous method such as the \textit{Bayesian model selection} described in Section~\ref{sec:Bayes} and used in \cite{Feroz2008cluster} for galaxy cluster detection. It consists in choosing between two contesting hypotheses based on the value of a parameter of the source model for which the prior distribution is known. Therefore we perform this comparison testing the source flux and applying a threshold to the flux instead of the SNR.
If $S_\mathrm{cut}$ is too low, then not all fake detections are removed. On the other side, if it is too high, then we may cut out all false positives but also begin to remove true ones, limiting our ability to detect low flux galaxies (see Fig.~\ref{fig:flux_cut_n_clusters} based on multiple tests on simulated source populations between 10 and 50 galaxies in the FOV). 

Similarly to \cite{Feroz2008cluster}, we define the following hypotheses:

\medskip
\noindent 
$H_0$: a galaxy with flux $S_\mathrm{min} < S \le S_\mathrm{lim}$ is contained in the field of view~$\Omega$,

\noindent
$H_1$: a galaxy with $S_\mathrm{lim} < S < S_\mathrm{max}$ is contained in $\Omega$,

\medskip
\noindent 
where $S_\mathrm{lim}$ is a lower bound of interest for the source flux.  
The model selection ratio given in equation~(\ref{eqn:bayes_factor}) is calculated as follows. 
For each hypothesis $H_k$, the associated evidence is:
\begin{equation}\label{eqn:evid_i}
\mathcal{Z}_k = \int \mathcal{L}(\mathbf{\Theta}) \pi _k(\mathbf{\Theta}) \mathrm{d}^N\mathbf{\Theta},
\end{equation}
where 
\begin{equation}
\pi _k(\mathbf{\Theta}) = \pi(l,m)\pi_k(S) \pi (\alpha) \pi (e),
\end{equation}
for $k=0,1$ are the priors defining the hypotheses. The source parameter priors are defined as in Section~\ref{sec:priors}, and they differ between the two hypotheses only for the flux: $\pi_0(S)$ and $\pi_1(S)$ are the power law distribution defined in equation (\ref{eq:flux}) normalised over the ranges $S_\mathrm{min} < S \le S_\mathrm{lim}$ and $S_\mathrm{lim} < S < S_\mathrm{max}$ respectively, and zero elsewhere.  
The corresponding local evidences are returned by \textsc{MultiNest} for each mode and we then cross-match the returned posterior modes between runs for each hypothesis by their position estimates using the software \textit{stilts}\footnote{\url{http://www.star.bris.ac.uk/~mbt/stilts/}}. We find that the $H_0$ run typically returns a much larger number of modes than for $H_1$. As F1 modes are removed by clustering first, this matching only involves F2 and true modes. The two hypotheses return different sets of F2 modes, but generally have at least one mode present at the expected position of a true mode, enabling us to associate the modes between the runs. The prior ratio Pr($H_1$)/Pr($H_0$) is obtained from the cumulative flux prior distribution: 
\begin{equation}\label{eqn:PHo}
\frac{\mathrm{Pr}(H_1)}{\mathrm{Pr}(H_0)} = \frac{\int_{S_\mathrm{lim}}^{S_\mathrm{max}} \pi(S)dS}{\int_{S_\mathrm{min}}^{S_\mathrm{lim}} \pi(S)dS} = \frac{S_\mathrm{max}^{-0.34}-S_\mathrm{lim}^{-0.34}}{S_\mathrm{lim}^{-0.34}-S_\mathrm{min}^{-0.34}}.  
\end{equation} 
We are thus able to compute the models' posteriors ratio $R$, and the probability of the mode being due to a detection of a galaxy with flux $S > S_\mathrm{lim}$ is:
\begin{equation}\label{eqn:probability_mode}
\mathrm{Pr}(H_1 \mid \mathbf{D})= R \cdot \mathrm{Pr}(H_0 \mid \mathbf{D}) = \frac{R}{1+R}.
\end{equation}

We compare performance of the flux cut and Bayesian approaches for detecting 50 galaxies within the FOV in two different cases. The tunable parameters of \textsc{MultiNest} are kept fixed between runs.

\begin{itemize}
\item [a)] \textit{Population with flux above 10~$\mu$Jy, corresponding to the typical weak lensing surveys threshold of SNR~$\ge 10$.}

We generate visibility data of a galaxy population with fluxes distributed according to equation~(\ref{eq:flux}) defined between 10-200$\mu$Jy. For the Bayesian modal selection, we ran \textsc{MultiNest} twice on the simulated visibility data: once with flux prior defined between 3-10$\mu$Jy and the second between 10-200$\mu$Jy, corresponding to $H_0$ and $H_1$ as above. Each run returns the local evidence for each posterior mode identified for the given hypothesis. After removing F1 modes through the clustering algorithm and matching mode detections between the two runs via position estimates, we use the local evidence values to compute the probability of each detection as belonging to a galaxy with flux greater than 10$\mu$Jy as described above. The final sample of galaxies is chosen by selecting modes with probability greater than 0.5. 
For 50 simulated galaxies, 47 modes are selected as true galaxies with no false positives.
An analysis using a single \textsc{MultiNest} run with flux prior ranging between 9-200~$\mu$Jy and applying a flux cut of~10$\mu$Jy, corresponding to the minimum source flux of our simulated catalog, led to all spurious being removed without the discarding of true positives and returned the same number of detected galaxies as the Bayesian selection. 
\medskip
\item [b)] \textit{Population with low flux range: 5-10~$\mu$Jy, corresponding to the typical galaxy catalog surveys threshold of SNR $\ge 5$.}

Visibility data of 50 galaxies is now simulated with fluxes distributed 
between 5-10$\mu$Jy. For implementing the Bayesian modal selection, we considered as two competing hypotheses a flux prior  defined between 3-5$\mu$Jy ($H_0$) and between 5-10$\mu$Jy ($H_1$). 
In this case, we find that the Bayesian selection returned 61 detections of which 41 true positives and 20 spurious whereas a run with flux prior down to 4~$\mu$Jy and a flux cut of 5~$\mu$Jy resulted in detecting 57 sources with 45 real and 12 spurious.
Further investigation in the low SNR regime is explored in Section~\ref{sec:low_snr_detection}.
\end{itemize}

We conclude that the threshold approach is the most efficient because it produces a lower fraction of false positives and is computationally cheaper as it only needs to run \textsc{MultiNest} once on the dataset.
We note that there is an option within \textsc{MultiNest} to only accept modes with evidence above a minimum  value provided by the user, but found that the flux cut approach described above was more robust to changes in the number of galaxies in the FOV and led to a lower number of false positives.

\begin{figure}
\centering
    \includegraphics[scale=0.6]{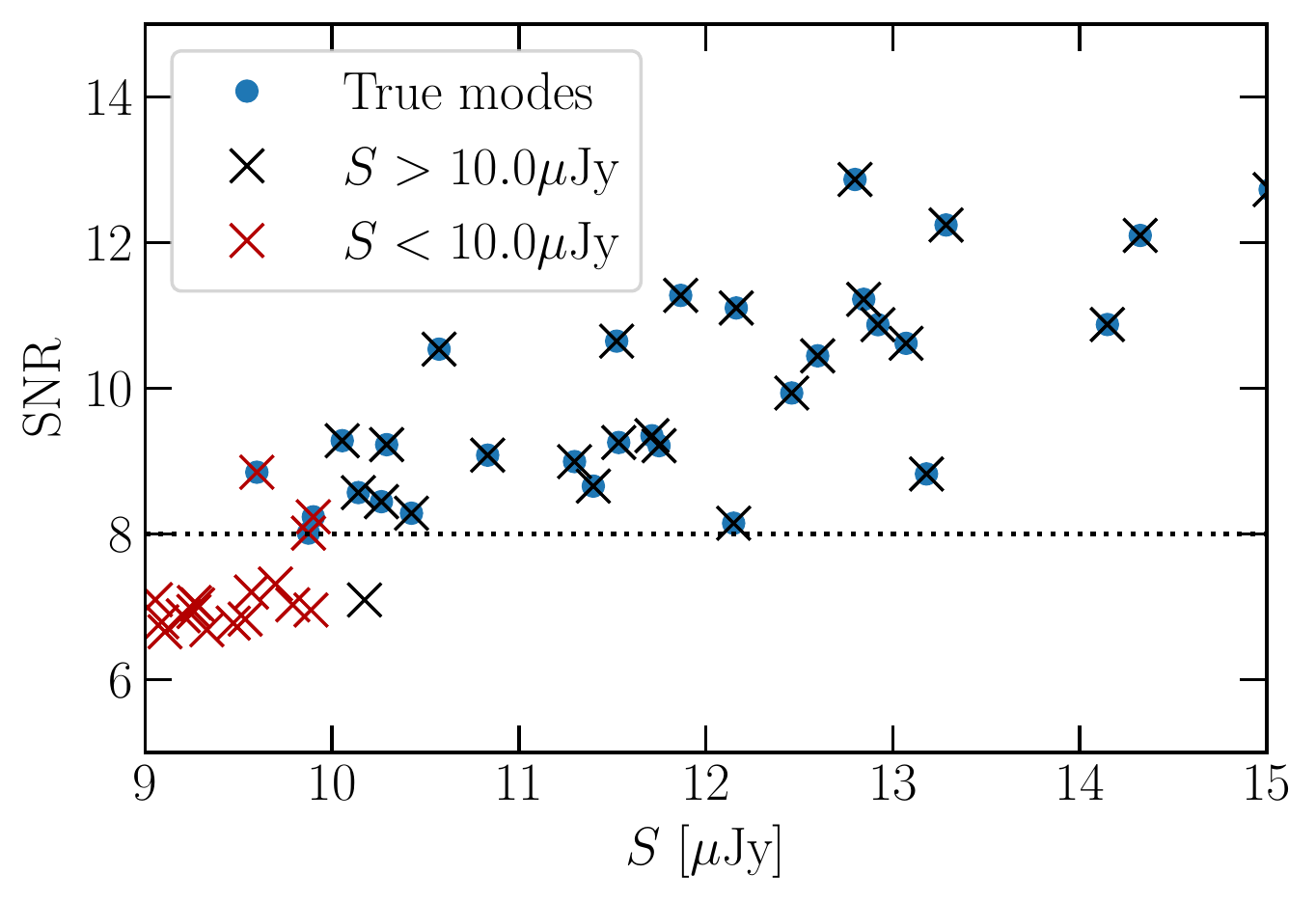}
    \caption{SNR versus estimated flux for all modes returned by \textsc{MultiNest} on a simulated population of 100 galaxies, with flux ranging between 10$\mu$Jy and 200$\mu$Jy. Blue circles show the `true' modes of the run (based on a cross-match with the input galaxy population); 98/100 are selected using a SNR threshold of 8 with no false positives. On the other hand, using $S_{\text{cut}}=10.0\mu$Jy returns 96/100 galaxies with 1 false positive. 
    Note that we have zoomed in to the low SNR and $S$ region of the plot to highlight the improvement of the SNR cut over a flux cut.}
    \label{fig:snr_vs_flux}
\end{figure}

Subsequent detections presented in this paper are obtained by applying a SNR cut as it takes into account not only the source flux but also the size of the source, improving the mode selection. For example, Fig.~\ref{fig:snr_vs_flux} shows that cutting at SNR = 8
recovers a slightly purer selection of the lowest SNR modes relative to applying a flux cut at 10~$\mu$Jy. We stress that our computation of the SNR is performed in the visibility domain as in \citet{rivi2016radiolensfit}, which usually returns a larger value than the standard peak brightness over the rms noise method used in the image domain (e.g. see Section 4.2.3 of \citet{rivi2018radio2}.

\subsection{Galaxy parameter estimates}

\begin{figure*}
	\centering
    \includegraphics[scale=0.63]{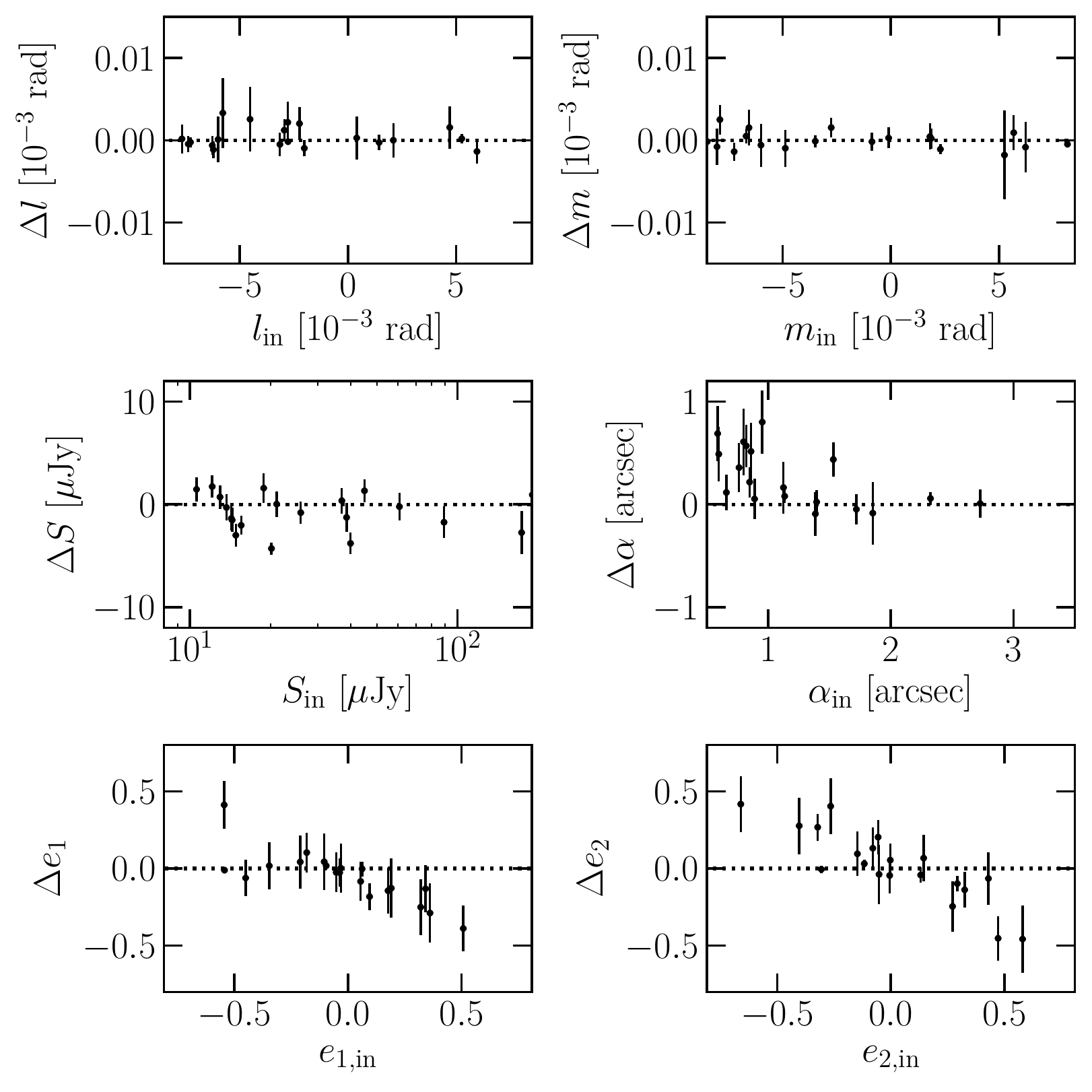}
    \caption{20/20 sources with input fluxes ranging between 10$\mu$Jy and 200$\mu$Jy (SNR~$\gtrsim 10$) distributed according to equation~(\ref{eq:flux}), detected using $N_\textrm{live}=8000$ points. For each detected galaxy, we plot the residuals of its estimated parameters (measured minus true value). As expected, position and flux measurements are much more accurate than scale-length and ellipticity components. Combined with estimates presented in figures~\ref{fig:50_param_estimates} and \ref{fig:100_param_estimates}, we find that position estimates appear free from systematic biases, whereas fluxes are generally underestimated. Smaller $\alpha$ values are overestimated, whilst the ellipticities are biased towards zero.}
    \label{fig:20_param_estimates}
\end{figure*}
\begin{figure*}
	\centering
    \includegraphics[scale=0.63]{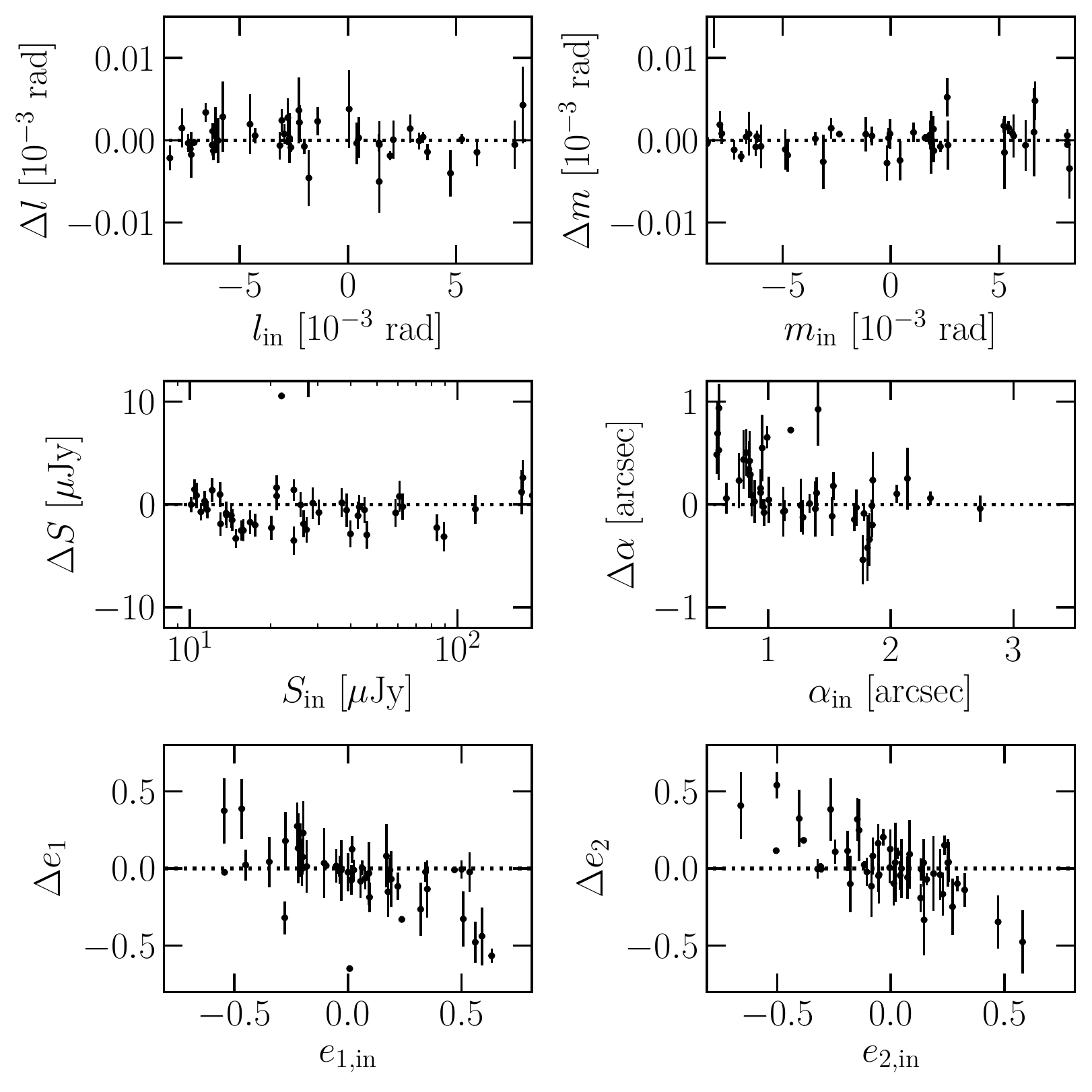}
    \caption{Residual plots for 47/50 sources with input fluxes ranging between 10$\mu$Jy and 200$\mu$Jy detected using $N_\textrm{live}=14000$ points. 6\% of the estimates for S, $\alpha$, e$_1$ and e$_2$ are inconsistent with the true value to within 5$\sigma$ and these also typically have very small uncertainties in their estimates.}
    \label{fig:50_param_estimates}
\end{figure*}
\begin{figure*}
	\centering
    \includegraphics[scale=0.63]{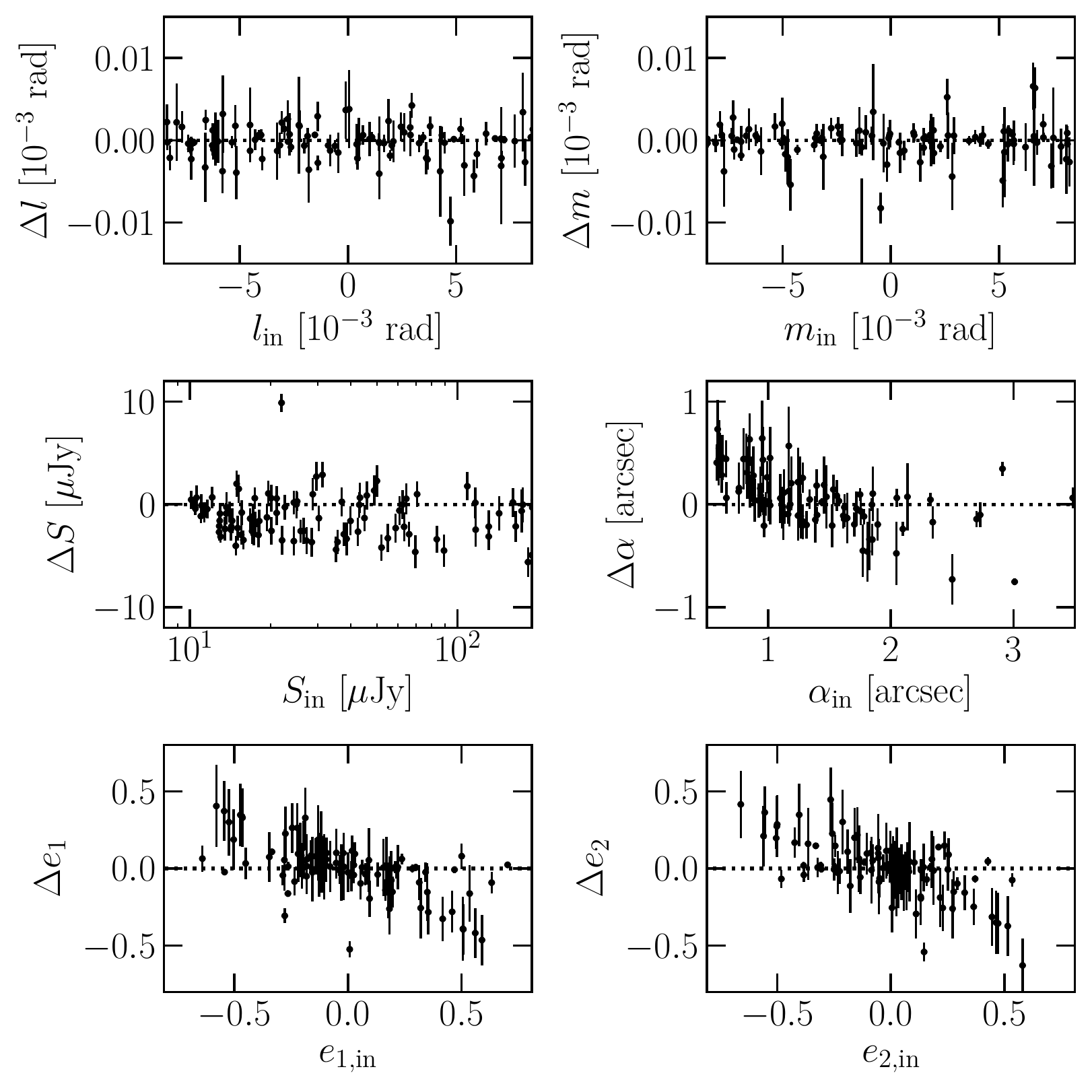}
    \caption{Residual plots for 98/100 sources with input fluxes ranging between 10$\mu$Jy and 200$\mu$Jy, detected using $N_\textrm{live}=35000$ points. About 3\% of the estimates for S, $\alpha$, e$_1$ and e$_2$ are inconsistent with the true value to within 5$\sigma$.}
    \label{fig:100_param_estimates}
\end{figure*}

We report our parameter estimates for each galaxy with the mean and standard deviation of the posterior samples returned for each posterior mode identified as `true' by our clustering and SNR cut selection algorithm. If needed, our approach also returns the full posterior distribution for each inferred source. Within a run, \textsc{MultiNest} typically convergences on the source positions earliest, before later converging on the shape parameters. If $N_{\textrm{live}}$ is not sufficiently high, then generally no convergence between runs is attained for $S$, $\alpha$, $e_1$ and $e_2$.

We show results obtained from simulated visibilities of populations in the radio weak lensing regime, i.e. with SNR~$\gtrsim 10$.
In Fig.~\ref{fig:20_param_estimates}, we plot our parameter estimates for the fitting of 20 galaxies within the field of view, using $N_\textrm{live}=8000$ live points.   
All sources are detected and with no false positives. We find that source positions and fluxes are fit accurately (although fluxes tend to be underestimated), while scale-lengths and ellipticity estimates deviate more from their true value (as expected, since these parameters are notoriously harder to fit). More specifically, scale-lengths tend to be overestimated for small input $\alpha$ values, whereas ellipticities are biased towards 0.  
We find similar behaviour in parameter estimates for the fitting of 50 galaxies (Fig.~\ref{fig:50_param_estimates}), but now with a source recovery rate of 94\% when using $N_\textrm{live}=14000$ (increased due to requiring a higher resolution at finding modes) and 100 galaxies (Fig.~\ref{fig:100_param_estimates}), where we recover 98\% of sources using $N_\textrm{live}=35000$. For the detected sources in the 20, 50 and 100 source simulations, the mean of the 1$\sigma$ uncertainty in position and flux parameter estimates is 0.4 arcsec and 1.2$\mu$Jy respectively. 
Under-estimation of source fluxes, which seems greater at larger values of the original flux, is probably due to the finite sampling of the uv coverage causing a loss of flux detection at short baselines, while over-estimation of the size for small sources may be due to the resolution of the telescope.

Relative to existing Bayesian methods in the visibility domain our approach suffers from lower accuracy in shape estimates even in this low source number density regime. This is mainly due to the usage of a single source model which does not account for interference between emission from other galaxies in the FOV. This is shown in \cite{rivi2018radio} where, although the source extraction method tries to remove as much as possible the contamination from the other sources, a neighbouring bias in the shear measurement is still estimated comparing to the case where there is a single source in the FOV. In \cite{rivi2018radio2} the multi-source model provides better shape measurement accuracy as expected. Also, we are measuring all the six galaxy parameters simultaneously, whereas such methods assume true values for source position and flux are known in the model fitting. 
As this method may be very useful for providing accurate source flux and positions, it could be combined with one of the above methods for a second fine-grained fit of the shape parameters. Shape estimates returned by \textsc{MultiNest} may be used to accelerate convergence of the HMC-based approach, or as the initial sky model for \textit{RadioLensfit}.

We now explore the ability of the method to detect sources with low SNR for constructing reliable galaxy catalogs for future SKA surveys. 
  
\subsection{Signal-to-noise detection threshold}
\label{sec:low_snr_detection}
\begin{figure}
	\centering
    \includegraphics[scale=0.45]{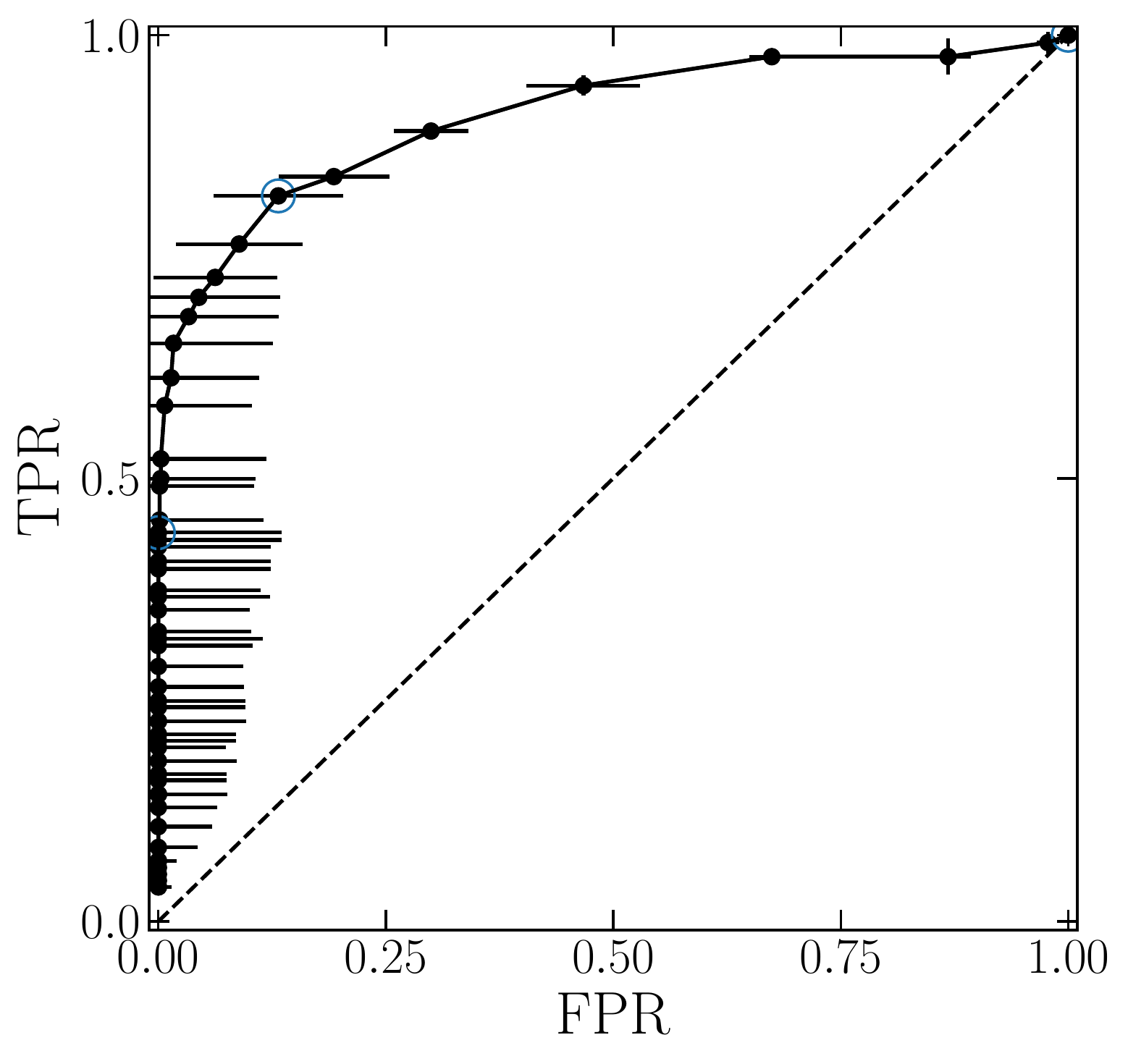}
   \caption{ROC plot for simulated sources within the 3-10$\mu$Jy range. Black markers are the mean value for a given SNR cut of FPR and TPR across three different 50 source population simulations, with error bars showing the range of FPR, TPR values at each SNR cut. Circled markers from left to right represent 
  SNR cuts in the visibility domain of $\sim$4.5, 3.3  and 2.6. Dashed line is $y=x$, which would be the expected curve for randomly guessing the mode type. The SNR threshold at which we have a good trade-off between TPR and FPR is at 3.3.}
    \label{fig:lowSNR_roc_plot_5_10}
\end{figure}

To characterize modal classification ability and to find the optimal signal-to-noise ratio threshold for detection, we construct a \textit{receiver operating characteristic} (ROC) curve (see Fig.~\ref{fig:lowSNR_roc_plot_5_10}), which is a plot of the true positive rate, TPR, defined as:
\begin{equation}
\label{eq:tpr}
\textrm{TPR}=\frac{N_\textrm{TP}}{N_\textrm{TP}+N_\textrm{FN}},
\end{equation}
against the false positive rate:
\begin{equation}
\label{eq:fpr}
\textrm{FPR}=\frac{N_\textrm{FP}}{N_\textrm{TN}+N_\textrm{FP}},
\end{equation} 
for different SNR threshold values, where $N_\textrm{TP}$, $N_\textrm{FP}$,  $N_\textrm{TN}$, $N_\textrm{FN}$ are the number of true positives, false positives, true negatives, and false negatives respectively, returned by our mode selection algorithm. 
The TPs, FPs, TNs and FNs are defined in the context of classifying a mode as real or spurious. We define the real modes as those which have their position estimates consistent with the position of a galaxy in the input source catalog, and spurious modes as those which do not have a match. Consistency is checked only for the positional parameters as these were found to be the most reliable and the easiest parameters to fit, while
the fitting is generally less consistent for fluxes and shape parameters.
After selecting the mode with the highest $\mathcal{Z}_{\textrm{loc}}$ in the cluster and performing the SNR cut, we have two sets of modes: those with SNR above and below the SNR$_{\textrm{cut}}$, which we define as the positives and negatives respectively. The true and false positives are modes above the SNR$_{\textrm{cut}}$ that correspond to real and spurious sources respectively, whilst the true and false negatives are modes below SNR$_{\textrm{cut}}$ that correspond to real and spurious sources respectively.

\begin{figure}
	\centering
    \includegraphics[scale=0.45]{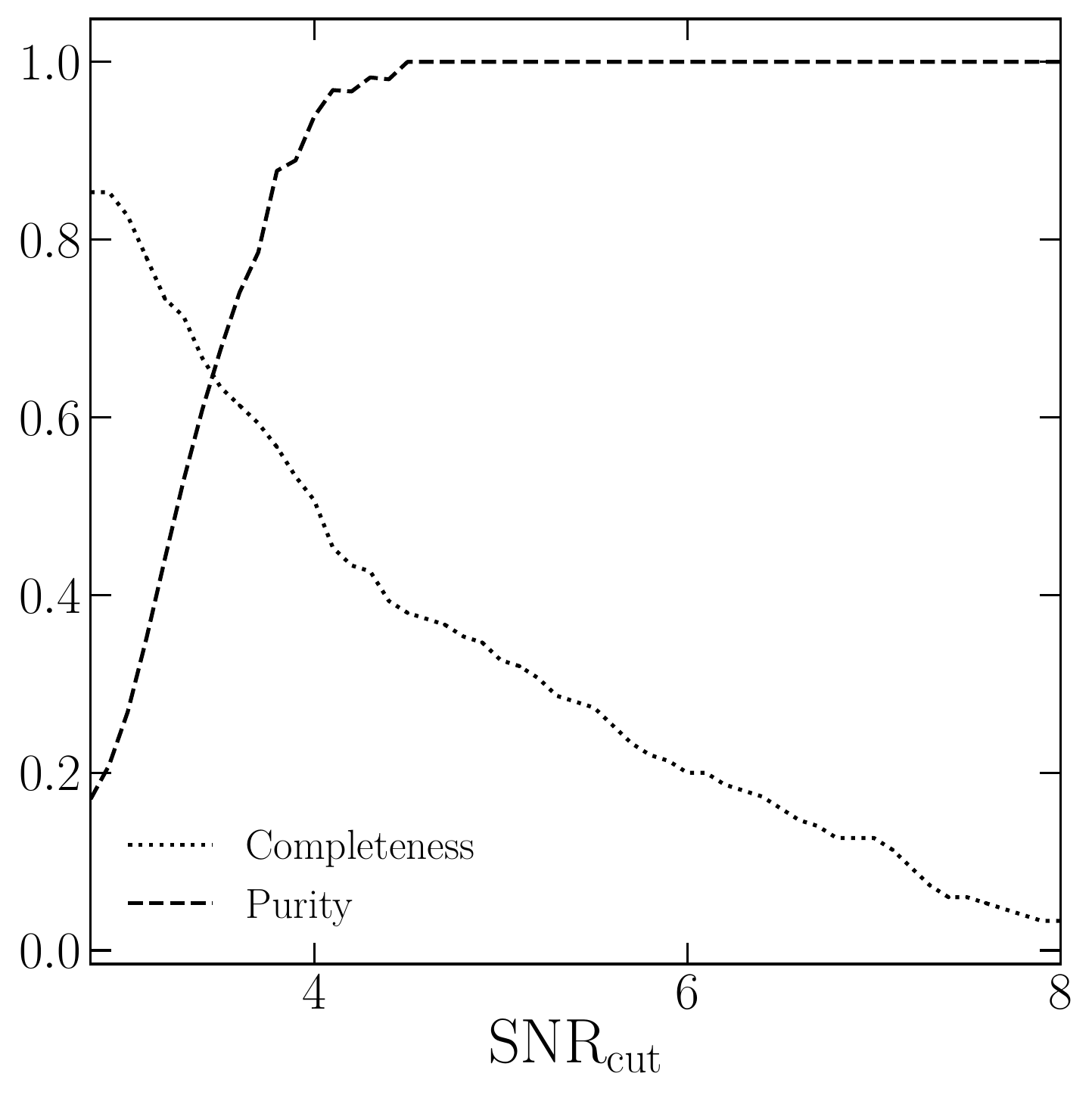}
   \caption{Completeness and purity plots corresponding to the ROC curve shown in Fig.~\ref {fig:lowSNR_roc_plot_5_10}. Purity increases with the SNR threshold but lowering the completeness of our final selection of galaxies. We find purity = 1 for SNR cuts above 4.5.}
    \label{fig:lowSNR_puritycompleteness_vs_scut}
\end{figure}

We simulate the visibilities of 50 galaxies within the FOV with fluxes limited to 3-10$\mu$Jy (i.e. SNR ranging between 3 and 13).
We set the domain of our flux prior to be defined between 3-200$\mu$Jy and its normalisation factor is adjusted accordingly.
The resulting ROC curve lies far from the line one would expect to obtain from random guessing modes as true or spurious (dashed line in Fig.~\ref{fig:lowSNR_roc_plot_5_10}) and the area under the curve (AUC) is 0.91, so our modal classifier still performs well on low SNR sources. 
The $\text{SNR}_\textrm{cut}$ at which we have a good trade-off between the TPR and FPR is at $\text{SNR}_\textrm{cut}\sim 3.3$ in the visibility domain; this is estimated as the value which maximises $\text{TPR}-\text{FPR}$.
To accompany the ROC curve, we also show in Fig.~\ref{fig:lowSNR_puritycompleteness_vs_scut} the \textit{purity} ($N_\textrm{TP}/(N_\textrm{TP}+N_\textrm{FP})$), and \textit{completeness} ($N_\textrm{TP}/N_\textrm{src}$) plot of our sample of galaxies after modal classification as a function of SNR$_\textrm{cut}$. We find that as before, if the threshold is too low, then our approach suffers from low purity but high completeness in the final modes returned. For example, when no selection is applied (at SNR$_{\text{cut}}\sim2.6$) then 16\% of the detected modes are true and only 85\% of the original population is detected, as we find the very low SNR sources are not detectable. As the threshold is raised, we increase purity but lower completeness. 
If high purity is preferred over completeness of the recovered galaxy population, then no false positives are generally found when applying SNR$_{\text{cut}} \sim 4.5$.


\begin{figure}
	\centering
    \includegraphics[scale=0.8]{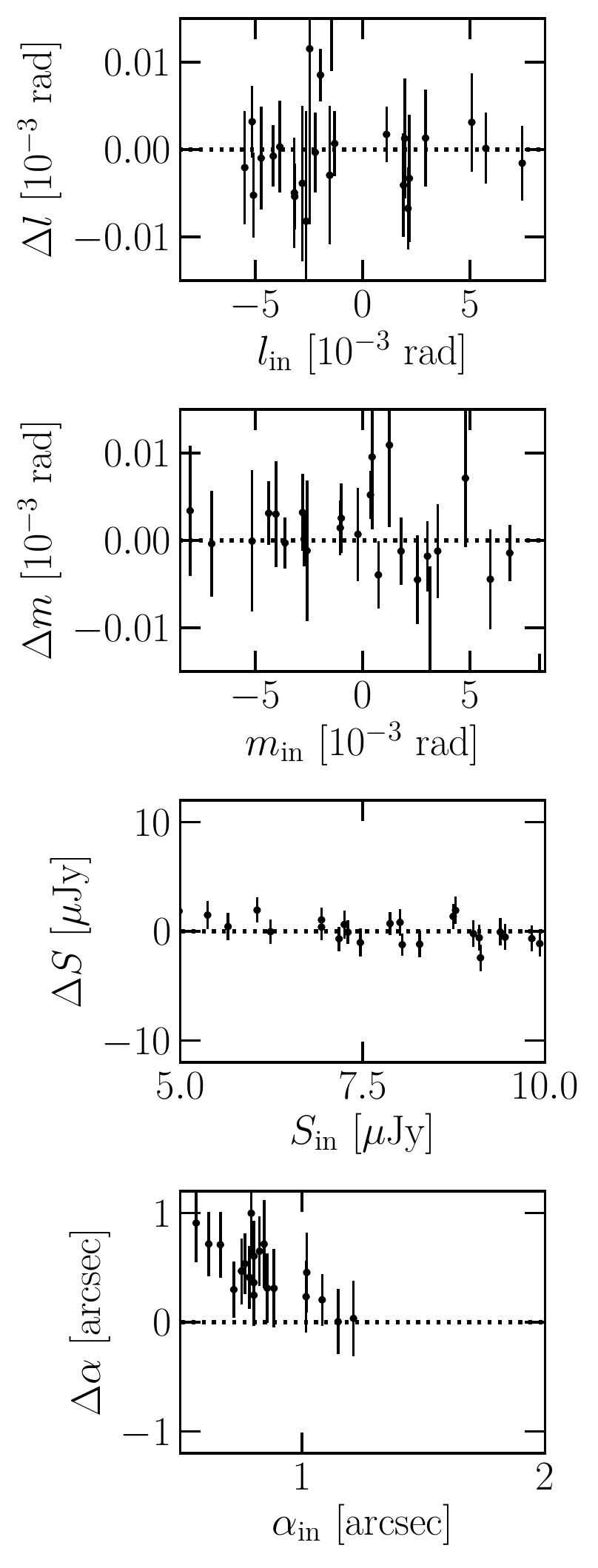}
   \caption{Parameter estimates for fitting of 50 sources in the FOV with SNR ranging approximately between 3 and 13. We use a flux prior down to $3\mu$Jy and select modes with SNR above 4.5 (the minimum SNR$_{\text{cut}}$ that produces no false positives), which returns 25/50 galaxies for this particular run, i.e. 58\% of the population with SNR~$\ge 5$.}
    \label{fig:lowSNR_param_estimates}
\end{figure}

We note that, as our method typically overestimates sizes in this regime, the SNR of true modes are usually lower than the SNR of the corresponding simulated sources. Therefore in order to recover most of the population above a given SNR, it may be required to apply a cut to the modes less than this SNR. For example, at SNR$_{\text{cut}} \sim 5$ we detect on average only $\sim$50\% of the simulated population with SNR~$\ge 5$, whereas lowering the SNR threshold would result in a greater recovery percentage of this population.

We show the galaxy parameter estimates of the detected sources of one simulation 
in~Fig.~\ref{fig:lowSNR_param_estimates}.
As expected for the low SNR regime, the ability to constrain the galaxy fluxes and size parameters is worse than with higher SNR sources, with the magnitude of galaxy sizes typically overestimated, but position measurements are still accurate. Across the recovered galaxies, the mean 1$\sigma$ uncertainty in position and flux parameters is 1.4 arcsec and 1.2$\mu$Jy respectively. 
Our method does not suffer any significant deterioration in the accuracy of recovered positional and flux estimates, although there can be a larger number of false positives in the low flux ranges.  

\section{Conclusions}
\label{sec:conclusions}
We have explored a novel Bayesian method for detecting galaxies from radio interferometric data using a single source model and sampling the resulting multimodal posterior with \textsc{MultiNest}. Without the need for any transformation to the image domain, our approach extracts out galaxies within visibility datasets, and estimates their positions and other properties from the appropriate posterior modes. 
We have tested this approach on simulated SKA1-MID observations of up to 100 star-forming galaxies in the field of view and searched for an acceptable SNR threshold where detections are reliable. 
To remove fake modes we propose to apply the mean shift clustering algorithm to group modes with similar estimated galaxy positions, setting the bandwidth for clustering to roughly the mean of the 1$\sigma$ uncertainty in positional parameter estimates (i.e. $\sim$0.4 arcsec). Once clustered, we select the mode with the highest evidence within each cluster as a true detection. The remaining false positives usually have low fluxes and large sizes, so may be removed with a suitable SNR cut. A more rigorous Bayesian hypothesis selection seems to be less efficient, besides the fact that it is much more computationally expensive.

We find that an estimated SNR threshold of $\sim 4.5$ is reasonable for mode selection as spurious modes should be a negligible fraction of the detected sources. From our tests, we also expect not to find spurious modes above SNR~$\sim 10$. We note that spurious modes which remain after performing a SNR cut could be identified in real datasets via cross-matching results with other surveys of the same field at different wave bands.  
Shape parameters fitting at SNR~$\gtrsim 10$ result to be less accurate than other methods proposed for radio weak lensing in the Fourier domain and using the same galaxy model \citep{rivi2018radio2,rivi2018radio}. This is expected because these methods take into account the signal interference between nearby galaxies by either using a multi-source model or removing an approximation of such source contamination. This is implemented assuming true flux and positions are well-known, whereas these parameters are free in our method. A further investigation for reducing this issue as well as estimating the reliability of this approach at large source densities, where galaxies may not be spatially well-separated, should be performed in future work. Moreover the impact of AGN structure must be studied as a non-negligible fraction of AGN population should be contained even in the faint radio sky. An initial discussion about this is presented in \cite{rivi2018radio2}.

Since our approach makes no assumptions about the number of galaxies within the field of view, it could become a useful tool for the development of accurate, reliable galaxy catalogs for the next generation of radio interferometers. 
In order to extend this method to larger source number densities similar to those expected for the SKA, and without using too large a number of \textsc{MultiNest} live points which would eventually lead to unfeasibly large computing times, a possible solution could be to split the pointing field of view into a number of tiles and then run the code on each one with positional prior limited to the space enclosed by each tile.
Once these improvements are made, this approach could also be used in conjunction with \textit{RadioLensfit} or HMC methods for a solely visibility domain-based analysis of galaxy populations for radio weak lensing.
 
As recent developments in image reconstruction techniques have been able to provide uncertainty quantification (see for example \cite{cai2018uncertainty,cai2018uncertainty2}), it would also be interesting to perform a detailed comparison of performance of our approach with these methods for detecting and characterising the observed galaxy population. 

\section*{Acknowledgements}
We thank Edward Edmondson for support with the UCL computer cluster, and Mike Hobson for usage advice on \textsc{MultiNest}. We also thank the anonymous reviewer for useful comments.

MR acknowledges the support of the Science and Technology Facilities Council via an SKA grant. 
FBA acknowledges the support of the Royal Society via a Royal Society URF award. This work is partially supported by EPSRS by grant EP/M011089/1, by STFC grant ST/M00113X/1, and by the Leverhulme Trust.


\bibliographystyle{mnras}
\bibliography{radio_mendeley}






\bsp	
\label{lastpage}
\end{document}